\documentclass{article}
\usepackage{amssymb}\usepackage{amsmath}\usepackage{epsfig}
\begin{document}
\title{Nonlinear supratransmission
\footnote{Journal of Physics Condensed Matter 15 (2003) 2933}}
\author{F. Geniet, J. Leon\\
Physique Math\'ematique et Th\'eorique, CNRS-UMR 5825\\
34095 MONTPELLIER  (FRANCE)}
\date{}\maketitle

\begin{abstract} A nonlinear system possessing a natural forbidden band gap can
transmit energy of a signal with a frequency in the gap, as recently shown for
the nonlinear chain of coupled pendula [Phys Rev Lett 89 (2002) 134102]. This
process of {\em nonlinear supratransmission}, occurring at a threshold exactly
predictable in many cases, is shown to have a simple experimental realization
with a mechanical chain of pendula coupled by a coil spring.  It is then
analyzed in more detail by first going to different (non-integrable) systems
which do sustain nonlinear supratransmission. Then a Josephson transmission
line (one dimensional array of short Josephson junctions coupled through
super-conducting wires), is shown to sustain also nonlinear supratransmission,
though being related to a different class of boundary conditions, and despite
the presence of damping, finiteness and discreteness. Finally the mechanism at
the origin of nonlinear supratransmission is a nonlinear instability, and it is
briefly discussed here.\end{abstract}

\section{Introduction}

It has been recently demonstrated that, in addition to energy spectral
localization (the Fermi-Pasta-Ulam recurrence phenomenon \cite{fpu}) and to
energy spatial localization (soliton generation from initial data
\cite{kruskal,zakshab,akns,solitons}), a nonlinear chain of oscillators possess
another striking fundamental property called {\em nonlinear supratransmission}
\cite{nous}. This phenomenon was shown to occur in the nonlinear sine-Gordon
chain, which possess a natural forbidden band gap, when it is submitted to
irradiation at a frequency in the stop gap.  While in a linear chain the signal
would exponentially vanish in the medium, it does not do so in the nonlinear
case if its amplitude exceeds a threshold value.  

In  \cite{nous} the irradiation of the medium was modelized by prescribing the
boundary value at one end of the chain. It is worth mentioning that nonlinear
transmission does occur also in the case of a true wave scattering, namely when
a monochromatic plane wave scatters onto a nonlinear medium with a frequency in
the gap \cite{jg-alex}. 

In the case of the sine-Gordon model, the threshold can be predicted exactly by
invoking the static one-breather solution \cite{nous}. In short, energy
penetrates the medium as soon as the amplitude $A$ of the harmonic driving, at
a frequency $\Omega$ in the gap, exceeds the maximum amplitude of the static
breather of frequency $\Omega$. This energy travels then through the medium by
means of nonlinear localized excitations (kinks, breathers, solitons).

The purpose of this paper is, after recalling results of \cite{nous}, to
display experimental results on the mechanical chain of pendula and to discuss 
the generality of our result by exploring extensions to different situations.
We shall in particular explore the robustness of nonlinear supratransmission
when the medium experiences damping, different nonlinearities and different
classes of boundary values.

The next section is devoted to a short reminder of the results published in
\cite{nous} and it is intended to settle formalism and basic facts about
nonlinear supratransmission. The model is the sine-Gordon chain submitted to a
Dirichlet condition at the origin on a vanishing initial background (pendula at
rest).  Then section \ref{sec:experiment} relates the experiment made with a
mechanical pendula chain, coupled by means of a coil spring, and which is
forced at one end by a periodic torque. A systematic exploration of the chain
response in a frequency range within the gap shows spectacular agreement with
the theory.  In section \ref{sec:charact} we numerically describe the
characteristics of the breathers generated in the sine-Gordon chain and
discover some simple relations between the parameters of the emitted breather
with respect to those of the boundary driving.  The section  \ref{sec:energy}
deals with the energy transmitted by the nonlinear medium as a function of the
driving amplitude at given frequency in the gap.  Particular emphasis will be
put on the effectiveness of the effect in different, not necessarily
integrable, cases.  The nonlinear instability which is the generating mechanism
of nonlinear supratransmission is briefly discussed in section
\ref{sec:instab}. Although the mathematical analysis is still to be constructed
we propose a quite simple illustration of the process by a perturbative
analysis of the sine-Gordon system driven close to a breather mode.

As another domain of study, we consider in section \ref{sec:josephson}  the
sine-Gordon model for Neumann conditions at the boundary (the derivative at the
origin is prescribed). This is a model for a chain of short Josephson junctions
whose first one is submitted to an external AC current. By using numerical
simulations, the process of nonlinear supratransmission is shown to hold, and
the threshold of energy transmission to obey a similar simple rule. 
Remarkably, in this case the energy flows by means of kink (or anti-kinks) and
not by breathers (or kink anti-kink pairs).  This is an interesting issue in
view of applications as kinks are the objects that have an experimental
signature (through the Josephson current).

\section{The sine-Gordon chain}\label{sec:sg}

\subsection{Generalities}

The following normalized discrete sine-Gordon chain of $N$ locally damped 
coupled  oscillators  $u_n(t)$
\begin{equation}\label{SG}
\ddot u_n-c^2(u_{n+1}-2u_n+u_{n-1})+\sin u_n=-\gamma(n)\dot u_n,
\quad n=1,\cdots , N\  ,\end{equation}
is submitted to the boundary value at the origin (Dirichlet condition)
\begin{equation}\label{forcing}
u_0(t)=A\sin\Omega t,
\end{equation}
acting on a medium initially at rest, namely
\begin{equation}\label{init}
u_n(0)=0,\quad  \dot u_n(0)=0.\end{equation}
In this section the damping coefficient $\gamma$ is used to model  a 
semi-infinite chain by an absorbing boundary. More precisely we take
$u_{N+1}(t)=0$ with
\begin{align}
&0\ge n\le m\ :\ \gamma(n)=0,\notag\\
&m<n\le N\ :\ \gamma(n)=a[1+\tanh(\frac{2n-m-N}{2b})],
\end{align}
where the parameter $b$ is adjusted to have a damping factor $\gamma(n)$
varying slowly from almost $0$ to almost $2a$ on the last $(N-m)$ particles. A
typical experiment will have e.g. $N=100$, $m=60$, $a=0.5$ and $b=3$. Note that
in section \ref{sec:josephson} we shall use instead the reflective boundary
condition $u_{N+1}=u_N$ for an homogeneous damping $\gamma(n)=cste$ for all $n$.

The equation \eqref{SG} is considered as a second-order ordinary differential
system for the  $N$ coupled oscillators $u_n(t)$. This system is then solved
with the subroutine {\tt dsolve} of {\tt MAPLE} software package which uses a
Fehlberg fourth-fifth order Runge-Kutta method.

\subsection{Forcing in the gap}

The linear dispersion relation $\omega(k)$ of the chain is
\begin{equation}\label{disp}
\omega^2 =1+2c^2(1-\cos k).
\end{equation}
 For a driving boundary \eqref{forcing} with frequency in the forbidden band 
gap (FBG), namely $\Omega<1$,  a linear chain would sustain the solution 
(evanescent wave) 
\begin{equation}\label{evanescent}
A\sin(\Omega t)\exp[-\lambda n].\end{equation}
The parameter $\lambda$ is given from the dispersion relation \eqref{disp}
written for $\omega=\Omega<1$ and $k=i\lambda$, namely
\begin{equation}\label{lambda}
\lambda={\rm arccosh}\left(1+\frac{1-\Omega^2}{2c^2}\right).\end{equation}

In the nonlinear case, in order to fit the boundary condition \eqref{forcing},
the {\em approximate} solution (exact in the continuous limit) is given 
instead by  the static breather
\begin{equation}\label{stat-breath}
u_b(n,t)=4\arctan\left[\frac{\lambda c\sin(\Omega t)}
{\Omega\cosh(\lambda(n-n_0))}\right],
\end{equation}
where the breather center $n_0$ solves
\begin{equation}\label{center}
A=4\arctan\left[\frac{\lambda c}
{\Omega\cosh(\lambda n_0)}\right].
\end{equation}
The above equation expresses that the  static breather fits the driving
field \eqref{forcing} by adjusting its position such as to match the
amplitude $A$. The spectrum of the breather signal does not match exactly
the monochromatic forcing \eqref{forcing} and it adapts by sending phonons
at third, fifth, etc..., harmonic frequencies.

\begin{figure}[ht]
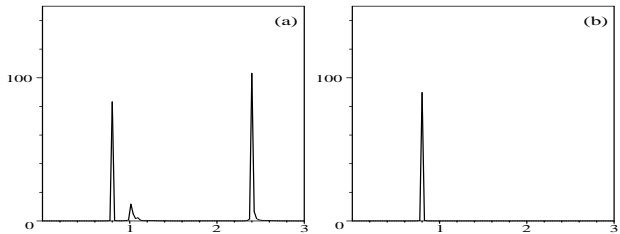

\centerline{\epsfig{file=fft-a.eps,height=3cm,width=4cm}
\epsfig{file=fft-b.eps,height=3cm,width=4cm}}
\caption{Spectra of the particle $50$ submitted to, (a) harmonic boundary 
driving at frequency $0.8$ and amplitude $2$ and to, (b) breather
boundary driving at frequency $0.8$. }
\label{fig:spectra-comp}\end{figure}

The above assertion is checked by performing two simulations of \eqref{SG}
at a given forcing frequency, say $\Omega=0.8$, and a given amplitude,
say $A=2$, in one case with the harmonic forcing \eqref{forcing}, in the
other case with the breather-like forcing $u_b(0,t)$ where the value of
$n_0$ is calculated from \eqref{center} with $A=2$. To avoid initial
shock we also set initial velocities matching the ones of the static breather.
Then we evaluate by Fourier transform the spectra of one particle of the chain
in both cases (we have selected the particle 50 on a chain of 100 pendula).  
The result is displayed on figure \ref{fig:spectra-comp} where
we see that phonons at frequency $3\Omega$ are indeed emitted for harmonic
forcing while no phonon appear for a breather-like forcing.
The small contribution at frequency $1$ in figure (a) represents a collective
motion of the pendula generated by the initial shock resulting from inadequation
of the initial velocities with the boundary value.

\subsection{Bifurcation threshold}

The adjustment of the breather center $n_0$ provided by \eqref{center} not
always has a solution. Indeed it works up to the maximum value $A=A_s$ of the
breather amplitude realized for $n_0=0$. Beyond this threshold, for a driving
boundary with $A>A_s$, we have shown in \cite{nous} that nonlinear
supratransmission occurs, i.e. the medium starts to transmit energy by means of
nonlinear modes generation (breathers and kink-antikink pairs). From
\eqref{stat-breath}, the threshold $A_s$ reads as the following function of the
frequency $\Omega$
\begin{equation}\label{threshold}
A_s=4\arctan\left[\frac{c}{\Omega}{\rm arccosh}\left(1+\frac{1-\Omega^2}{2c^2}
\right)\right],\end{equation}
which has the approximate value
\begin{equation}\label{thres-approx}
A_s\sim 4\arctan[\frac{\sqrt{1-\Omega^2}}\Omega].\end{equation}
which would hold for the continuous sine-Gordon equation obtained in the
limit $1/c\to0$. Note that the maximum difference (when $\Omega$ varies
in $[0,1]$) between the above two 
expressions is already $1\%$ for $c=1$ and goes down to  $0.01\%$ for $c=10$, 
the value we have used in most of the numerical simulations.

\begin{figure}[ht]
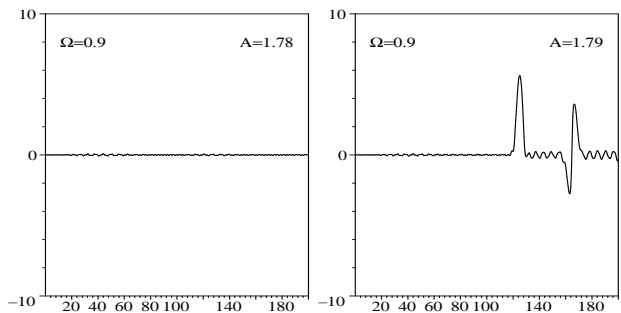

\centerline{\epsfig{file=noth.eps,height=4cm,width=4cm}
            \epsfig{file=breath.eps,height=4cm,width=4cm}}
\caption{Representation of $u_n(t)$ as a function of time for $n=60$ in the 
case $\Omega=0.90$ for two amplitudes.}\label{fig:motions}\end{figure}
This qualitative definition of a bifurcation threshold can be checked on
numerical simulations of \eqref{SG} with the boundary condition \eqref{forcing}
by varying, at given frequency $\Omega$, the amplitude $A$ around the above
value $A_s=1.803$ (for $c=4$). There are many means to determine appearance of
nonlinear supratransmission, a simple one being the observation of the motion
of one particle of the chain. As an example we display in figure
\ref{fig:motions} the motion of the particle $60$ of a chain of  $200$
particles driven at frequency $0.9$ at amplitudes $A=1.78$ (no
supratransmission)  and $A=1.79$ (supratransmission) for a coupling factor
$c^2=16$.  Each large oscillation in the second figure corresponds to a
breather passing by. Two of them are generated and cross the site $60$ at times
$120$ and $160$.  The small oscillation seen between the humps are the harmonic
phonons, mainly of frequency $3\Omega$.

\begin{figure}[ht]
\centerline{\epsfig{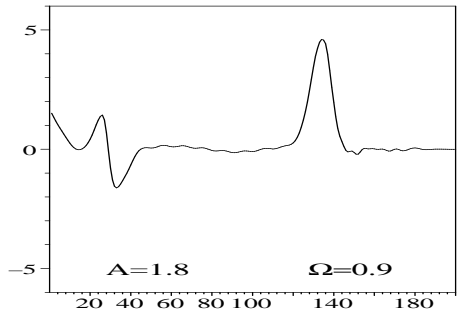}}
\caption{Plot of $u_n(t)$ as a function of $n$ at $t=120$.}
\label{fig:actual}\end{figure}
For illustration we show on figure \ref{fig:actual} a picture of oscillators
amplitude $u_n(t)$ at given time (here $t=120$) in the case of nonlinear
supratransmission obtained for $A=1.8$ and $\Omega=0.9$. A first breather 
propagates to the right while a second one is just being generated near the
origin. Note the amplitude of the breather with respect to the one of the
driving boundary.

\begin{figure}[ht]
\centerline{\epsfig{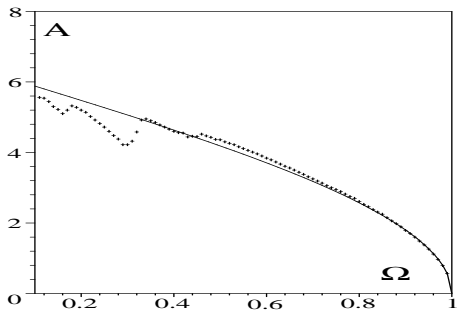}}
\caption{ Bifurcation diagram in the $(A,\Omega)$ plane. The solid line is the 
threshold expression \eqref{threshold}. 
The crosses indicate the lowest value of $A$ 
for which nonlinear supratransmission is seen on numerical simulations.}
\label{fig:bif}\end{figure}

By a systematic exploration of the chain response we draw figure \ref{fig:bif}
obtained for $200$ particles with a coupling $c^2=100$ (some experiments have
been actually made with smaller coupling and less number of points to shorten
computation times) for a typical time of $200$ (for frequencies close to the
gap value $1$, time had to be increased up to $500$). The points on figure
\ref{fig:bif} are obtained with an absolute precision of $10^{-2}$ for the
amplitude $A$. They are compared to the theoretical threshold expression
\eqref{threshold} (continuous curve).

The figure \ref{fig:bif} shows excellent agreement to formula \eqref{threshold}
apart for some discrepancies starting below $0.33$ and $0.18$. This results
from the driving which, thanks to the nonlinearity, generates phonons at
multiple frequencies (here third, fifth, etc...). If these frequencies lie in
the phonon band, the phonons move away from the boundary and have no effect on
the forcing. If however they lie in the FBG, the related phonons do not
propagate (which we call phonon quenching) and stick on the boundary where they
add contribution to the driving.  This effect indeed disappears when driving
the system \eqref{SG}
with the exact breather boundary value $u_b(0,t)$ for which we
have checked that nonlinear supratransmission {\em never occurs} at an
amplitude $A<A_s$, while it occurs already at $A=A_s$ (or for very small
deformation of the perfect breather).

\section{Experiments on a mechanical chain}\label{sec:experiment}

\begin{figure}[ht]
\centerline{\epsfig{file=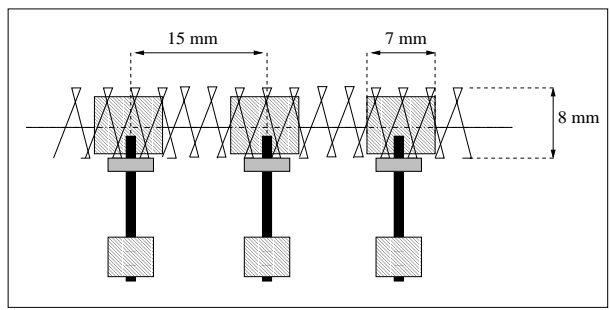,height=4cm,width=8cm}}
\caption{Sketch of the mechanical chain.}\label{fig:chaine}\end{figure}

Our purpose here is to show that the process of nonlinear supratransmission can
be easily realized experimentally on a chain of coupled pendula  as depicted in
figure \ref{fig:chaine}.  This chain has been built following M. Remoissenet
\cite{solitons}.  The pendula rotate freely around a piano wire stretched
between two supports, they are coupled together by a coil spring tightened to
each pendulum by a screw.

Such a chain of 48 pendula is driven by an electrical engine steered by a
generator of sinusoidal tension. Upon varying the frequency, at low amplitude,
in the phonon band, we can determine the parameters of the chain (coupling
constant $\sigma^2$ and angular eigenfrequency $\nu_0$) by measuring the
wavelength $2\pi/k$ of produced wave (in units of pendula number).  
By comparison to the dispersion relation of the linear chain
\begin{equation}\label{disp-mech}
\nu^2 =\nu_0^2+2\sigma^2(1-\cos k).
\end{equation}
we determined the following parameter values
\begin{equation}\label{param-mech}
\nu_0 =15\ {\rm Hz}\ ,\quad \sigma=32\ {\rm Hz}
\end{equation}
Then in a time normalized to the eigenfrequency, we have the dispersion relation
\eqref{disp} with
\begin{equation}\label{couplage}
c=\frac{\sigma}{\nu_0}=2.13,\end{equation}
which is the fundamental parameter of the model, entering in particular
expression \eqref{threshold} of the threshold amplitude for nonlinear
supratransmission. This is the formula we want here to confront to experiments.

\begin{figure}[ht]
\centerline{\epsfig{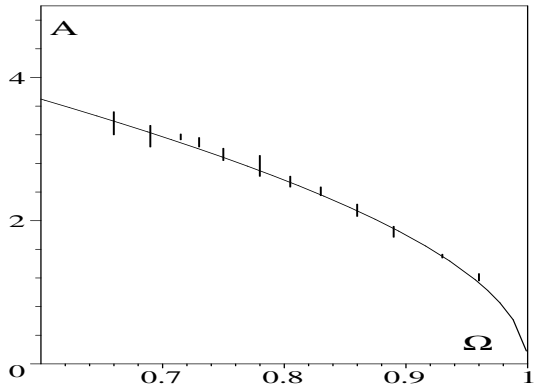}}
\caption{Experimental values of nonlinear supratransmission threshold
compared to expression \eqref{threshold}.}
\label{fig:exp-res}\end{figure}

We proceed with a systematic
exploration of the chain response to a signal frequency in the gap (of value
less than $\nu_0$). The method consist in increasing slowly, at fixed
frequency, the amplitude of the driving, up to the time when a nonlinear mode
is seen to be generated (a picture of a breather generated at a frequency of
$0.85$ in normalized units is displayed in \cite{nous}). Repeating 3 times, for
each driving frequency value, the measurements of the driving amplitude that
generates a nonlinear mode, we eventually obtain the figure \ref{fig:exp-res}
which displays the measured threshold amplitude in terms of the signal
frequency (normalized). There, the full line curve is the function
$A_s(\Omega)$ given in \eqref{threshold} with $c=2.13$.

Despite the small number of pendula, inducing reflection at the open end,
the inherent damping, and other mechanical imperfections, the experiments
provide a spectacular realization of the theoretical threshold prediction.

\section{Characteristics of the emitted breathers}\label{sec:charact}

One important issue concerns the nature and the characteristics of the
nonlinear modes which propagate in the medium. In the first instance, we have
to check that the emitted structures correctly match the moving breather,
at velocity $v<1$, which is  convenient to write as 
\begin{equation}\label{general-breath}
u_{v}(n,t)=4\arctan\left[
\frac{r\sin\left(\frac{1}{\sqrt{1+r^2}} \frac{t-vn/c}{\sqrt{1-v^2}}\right)}
{\cosh \left(\frac{r}{\sqrt{1+r^2}} \frac{n/c-vt)}{\sqrt{1-v^2}}\right)} 
\right]\ .
\end{equation}
By choosing correctly the two parameters $r$ and $v$ which represent
respectively the amplitude and the group velocity of the breather, together
with the space and time origins of the solution $u_{v}(n-n_0,t-t_0)$, one can
match it to the asymptotic numerical solution with high precision.  This
enables us to identify correctly the emitted breathers, to determine their
characteristic with a great accuracy, and to observe their possible
disintegration into kink-antikink pairs.

\begin{figure}[ht]
\centerline{\epsfig{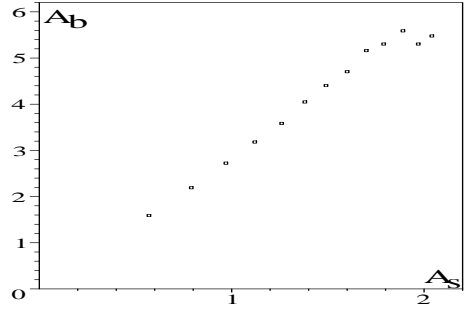}}
\caption{Amplitude of the emitted breathers in terms of the amplitude of 
excitation, just above the emission threshold.}
\label{fig:ampl-breath}\end{figure}

\begin{figure}[ht]
\centerline{\epsfig{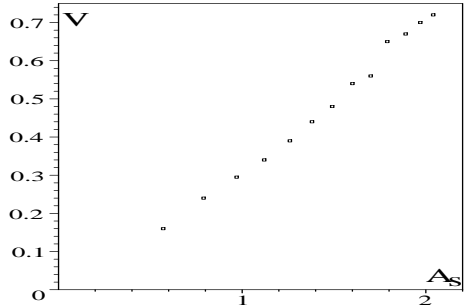}}
\caption{Group velocity of the emitted breathers,
as a function of the amplitude of excitation, just above the emission 
threshold.}\label{fig:vit-breath}\end{figure}

The results of such estimations are displayed in figure \ref{fig:ampl-breath}
obtained as follows. For each driving frequency $\Omega$, we have driven the
system at threshold amplitude $A_s(\Omega)$ and then determined at the the
parameters $\{r,v\}$ of the emitted breathers (a given experiment at threshold
driving produces repeatedly the same breather).  Then the amplitude of the
propagating breather  is plotted as a function of the amplitude of excitation
$A_s(\Omega)$  for various frequencies. Similarly, figure \ref{fig:vit-breath}
shows the group velocity of the breather as a function of $A_s(\Omega)$.

Both figures \ref{fig:ampl-breath} and \ref{fig:vit-breath} correspond to a
region of the driving frequency close to the gap $0.87 \leq \Omega \leq 1$. 
For lower frequencies the emitted breathers, when they occur, are unstable and
decay into kink-antikink pairs (this will be discussed later).  The main
conclusion here is the existence of a linear relation between the amplitude
$A_b$ (and the velocity $v$) of the generated breather and the driving
amplitude at threshold $A_s$. So far we have no theoretical interpretation
of these observations.

\begin{figure}[ht]
\centerline{\epsfig{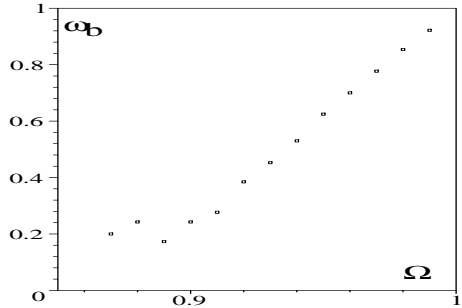}}
\caption{Proper frequency of the emitted breathers,
as a function of the driving frequency $\Omega$, just above the emission 
threshold.}\label{fig:freq-breath}\end{figure}

Another characteristic of the breather is its proper
frequency $\omega_b$, related to the apparent pulsation
period $T$ by the usual (relativistic) relation
\begin{equation}
\omega_{b}=\frac1{\sqrt{1+r^2}}=\frac{2 \pi }{T\sqrt{1-v^2}}\ .
\end{equation}
This frequency is plotted in figure \ref{fig:freq-breath}
as a function of the driving frequency $\Omega$ and,
again, we note a non trivial linear dependence.

We now turn to the stability of the emitted breathers which, for $\Omega \leq
0.88$  decay into kink-antikink pairs. This can be qualitatively described by
studying the binding energy $W$, given by 
\begin{equation} \label{binding-E}
W = 2 E_k - E_b = 
\frac{16}{\sqrt{1-v^2}} \ \left( 1-\sqrt{\frac{r^2}{1+r^2}} \right)\ ,
\end{equation}
where $E_k$ is the single kink energy and $E_b$ the breather energy
\cite{solitons}.
\begin{figure}[ht]
\centerline{\epsfig{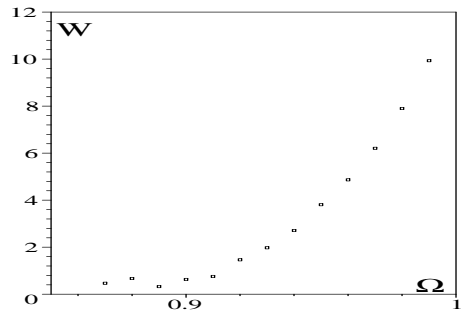}}
\caption{Binding energy of the emitted breathers,
as a function of the frequency of excitation, 
just above the emission threshold.}
\label{fig:energ-breath}\end{figure}

We have evaluated the binding energy $W$ of the breathers produced in the
simulations (by plugging in (\ref{binding-E}) the measured values of $r$ and
$v$), the result of which is plotted in figure \ref{fig:energ-breath} in terms
of the frequency $\Omega$ of the driving boundary. This figure shows that the
binding energy of the breather decreases with the driving frequency, and goes
to zero for $\Omega \lesssim 0.88$.  Thus the breather decay  in the frequency
range $[0,0.8]$ is well understood from formula (\ref{binding-E}).

\section{Energy transmission}\label{sec:energy}

\subsection{Basic expression}

Nonlinear supratransmission is a process where large amount of energy  flows
through the medium. Our purpose here is to evaluate numerically this energy
for amplitudes around the threshold value. The theoretical expression of the
energy flow is calculated hereafter for a generic nonlinearity deriving from
a potential energy $V(u_n)$ (the sine-Gordon case 
\eqref{SG} corresponds to $V(u_n)=1-\cos u_n$).

From the energy density
\begin{equation}\label{hamilt}
H_n=\frac12\dot u_n^2+\frac{c^2}2(u_{n+1}-u_n)^2+V(u_n),\quad
n=1,\cdots \infty\ ,\end{equation}
and the evolution equation  follow the conservation law
\begin{equation}\label{cons-law}
\frac{\partial}{\partial t}H_n+(J_{n+1}-J_n)=0,\end{equation}
with the current 
\begin{equation}
 J_n=-c^2\dot u_n(u_n-u_{n-1}).\end{equation}
Incorporating the potential energy resulting from the coupling of the
first particle $u_1$ to the boundary $u_0$, the {\em total energy} of the 
system reads
\begin{equation}\label{ener-tot}
E=\sum_{n=1}^{\infty}H_n+\frac{c^2}2(u_1-u_0)^2.\end{equation}
In our case $u_0(t)$ is the driving \eqref{forcing} and the chain is supposed
infinite with $u_n(t)\to0$ as $n\to\infty$.  

Upon time derivation, with help of the conservation law, and using the
assumed asymptotic $J_n\to0$ for large $n$, we arrive eventually at
\begin{equation}
\frac{\partial}{\partial t}E=c^2\dot u_0(u_0-u_1).\end{equation}
Hence the total energy injected in the medium during time $T$ reads
\begin{equation}\label{energy}
E=c^2\int_0^Tdt\ \dot u_0(t)[u_0(t)-u_1(t)].\end{equation}
Choosing for $T$ an integer multiple of the period of excitation makes this
energy to vanish identically in the linear case if the driving frequency falls
in the FBG.
  
\subsection{Numerical simulations}

\begin{figure}[ht]
\centerline{\epsfig{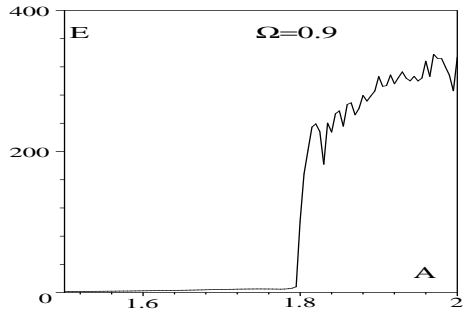}}
\caption{Energy $E$ injected in the sine-Gordon nonlinear chain 
for $T=140$ as a function of the driving amplitude $A$.}
\label{fig:sg-09}\end{figure}

In the nonlinear case, expression \eqref{energy} is computed  numerically on a
chain of 60 particles with a coupling parameter $c^2=16$ and an absorbing end
working over the last 40 particles.  For a driving frequency $0.9$ and
amplitudes running from $1.5$ to $2.0$, we obtain the  figure \ref{fig:sg-09}
where the bifurcation is seen to occur for $A=1.80$,  the value predicted by
formula \eqref{threshold}. This simulation has been run for frequencies in
the range $[0.2,0.9]$, with comparable expected results.

Our approach stems from the existence of a breather solution of the model
equation, allowing to determine the threshold amplitude.  It is then worth
wondering if the process is robust {\em against} non-integrability.  

To give a partial still instructive answer to this question, we have performed 
numerical simulations of two non-integrable evolution in the same class,
i.e. with Hamiltonian \eqref{hamilt}.

\begin{figure}[ht]
\centerline{\epsfig{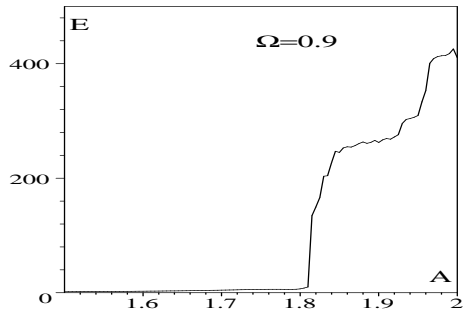}}
\caption{Energy in the Klein-Gordon chain \eqref{KG}.}
\label{fig:kg-09}\end{figure}

First the Taylor truncated expansion of sine-Gordon (fifth order
is kept to ensure a confining potential at large $u_n$) reads as the
nonlinear Klein-Gordon chain:
\begin{equation}\label{KG}
\ddot u_n-c^2(u_{n+1}-2u_n+u_{n-1})+ u_n-\frac1{3!}u_n^3+
\frac1{5!}u_n^5=0,\quad n=1,\cdots, N\ .\end{equation}
This system is solved with the boundary driving (\ref{forcing}) and the
energy (\ref{energy}) is computed for the same parameter values as for figure
\ref{fig:sg-09}. The result is displayed on figure \ref{fig:kg-09}. 

By scanning the frequency range in the gap, we have obtained that the process
occurs down to $\Omega=0.7$ and then disappears. We suspect that for such a
polynomial potential energy, at high forcing amplitude, the {\em incoming wave}
sees an almost parabolic potential, while for low amplitude driving the {\em
incoming wave} does feel the actual structure of the local potential.

\begin{figure}[ht]
\centerline{\epsfig{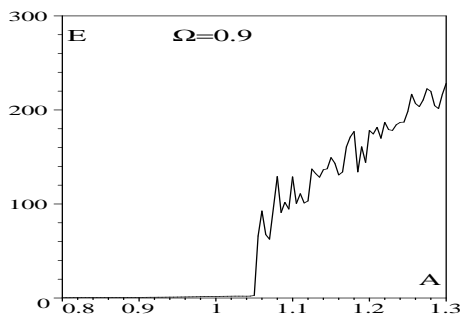}}
\caption{Energy in the double sine-Gordon chain \eqref{DSG}.}
\label{fig:dsg-09}\end{figure}

Another interesting non integrable Hamiltonian evolution where the local
potential has a periodic structure is the double sine-Gordon chain
\begin{equation}\label{DSG}
\ddot u_n-c^2(u_{n+1}-2u_n+u_{n-1})+ \frac13[\sin u_n+\sin 2u_n]=0
,\quad n=1,\cdots, N\ .\end{equation}
Once again we have solved this system in the same situation as before
and obtained the brutal energy flow of figure \ref{fig:dsg-09}.
In this case the process holds for any frequency (tested from $\Omega=0.2$)
just as in the sine-Gordon case.

Thus we have seen that in those two cases nonlinear supratransmission  
does work which is a strong indication that it is a generic nonlinear
process. An interesting question is the mechanism that generates nonlinear 
supratransmission, which is discussed now.

\section{Generating instability}\label{sec:instab}

The process at the origin of nonlinear supratransmission is a
nonlinear instability. Although it is not yet fully understood, we discuss
here some of its relevant aspects. 

To that end we explore the properties of the sine-Gordon chain \eqref{SG}
submitted to initial-boundary value data that precisely mimic the
breather $b_n(t)$ centered at $n_0=0$ as defined in \eqref{stat-breath} and
which is convenient to write as 
\begin{align}
&b_n(t)=4\arctan\phi_n(t)\ ,\label{breath}\\ 
&\phi_n(t)=\frac{\lambda c}\Omega\frac{\sin\Omega t}{\cosh\lambda n}\ .
\label{phi-breath}\end{align}
Note that this is an exact solution in the continuous
limit $1/c\ll 1$ and $\{x=n/c,\ \lambda c=\kappa,\ \Omega^2=1-\kappa^2\}$.

In order to study the behavior at the threshold, we impose the boundary value
\begin{equation}\label{bound-b}
 u_0(t)=(1+\epsilon)b_0(t)\ ,\end{equation}
together with compatible initial data
\begin{align}\label{init-b}
&u_n(0)=(1+\epsilon)b_n(0)\ ,\nonumber\\
&\dot u_n(0)=(1+\epsilon)\dot b_n(0)\ .
\end{align}
The boundary value at the other end  of the chain ($n=N$) can be taken
as an absorbing end to simulate the infinite line or as the breather value 
$b_N(t)$, e.g. to check the accuracy of the solution.

The parameter $\epsilon$ measures the departure from the exact solution. We are
of course interested in what happens when $\epsilon$ is positive. For
$\epsilon=0$ we simply generate the approximate solution $b_n(t)$ which in the
numerical simulations is  marginaly stable (and would be indeed stable for a
breather centered in $n_0<0$).

In the case $\epsilon>0$, the numerical simulations of the sine-Gordon model
immediately generate nonlinear supratransmission. From the initial-boundary
value problem, it is natural to seek a solution as a perturbation of the
breather under the form
\begin{equation}
u_n(t)=b_n(t) +\epsilon\eta_n(t)\ ,\end{equation}
which by \eqref{SG} obeys at order 1 in $\epsilon$
\begin{equation}\label{eq-eta}
\ddot \eta_n-c^2(\eta_{n+1}-2\eta_n+\eta_{n-1})+C_n\eta_n=
\epsilon D_n\eta_n^2\ .\end{equation}
The variable coefficients $C_n(t)$ and $D_n(t)$ of this equation are given by
\begin{align}\label{pot-inst}
&C_n=\cos b_n=1-\frac{8\phi_n^2}{(1+\phi_n^2)^2}\ ,\nonumber\\
&D_n=\frac12\sin b_n=2\phi_n\frac{1-\phi_n^2}{(1+\phi_n^2)^2}\ .
\end{align}

\begin{figure}[ht]
\centerline{\epsfig{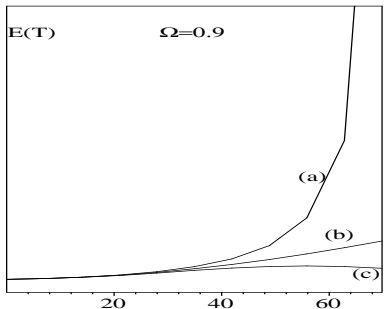}}
\caption{Energy $E(T)$ given by \eqref{energy} computed for the chain 
\eqref{eq-eta}. (a) exponential growth for $\epsilon=0.005$, (b)  
linear growth for $\epsilon=0$ and 
(c) oscillations for $\epsilon=-0.005$. 
$E(T)$ varies here on the scale $[0,2\ 10^4]$.}
\label{fig:lin}\end{figure}

The initial-boundary value problem that goes with \eqref{eq-eta} can be taken
to results from \eqref{bound-b}, namely
\begin{equation}
\eta_0(t)=b_0(t)\ ,
\end{equation}
with the related initial data. Then we observe the following fundamental
facts: 
\begin{itemize}
\item for $\epsilon<0$ the system \eqref{eq-eta} is stable showing a long 
period oscillatory behavior,
\item for $\epsilon>0$ the system is unstable showing an exponential growth
of the oscillations.
\end{itemize}
These observations are illustrated by figure \ref{fig:lin} where we have
computed the energy transmitted by the chain \eqref{eq-eta} according to
formula \eqref{energy}, as a function of time (the points in time are
chosen as integer multiples of the period $2\pi/\Omega$).
The parameters related to this experiments are  $N=100$,  $c=10$ and the
energy has been calculated for 10 points (up to $t=20\pi/\Omega$). 
We have also displayed results for the {\em linear version} of system
\eqref{eq-eta} (read with $\epsilon=0$) which is marginaly unstable showing a
linear growth of the oscillations. The exponential growth in the case
$\epsilon>0$ is the signature of the instability which is the mechanism of
nonlinear supratransmission. 
The mathematical approach of this instability for \eqref{eq-eta} is reported to
future studies.

\section{Josephson junctions array}\label{sec:josephson}

\subsection{The model}

A Josephson junction behaves just like a single pendulum, the rotation
amplitude being replaced by the phase difference between wave functions of
Cooper pairs \cite{junction}. By connecting the Junctions in parallel with
super-conducting wires, the resulting model equation is just the sine-Gordon
discrete system submitted to damping and constant torque \cite{array}
\begin{equation}\label{JSG}
\ddot u_n+\gamma\dot u_n+\sin u_n=
J+c^2(u_{n+1}-2u_n+u_{n-1}),\quad n=1,\cdots, N\ .\end{equation}
Here above $\gamma$ is the constant damping along the array, $J$ is the
normalized intensity of the applied current and time has been normalized to the
plasma frequency.

We consider here only the one-dimensional geometry but it is worth mentioning
the coupled arrays (ladders) who revealed as a means to generate discrete
breathers by convenient initial conditions \cite{roto-breath} with subsequent
striking experimental realizations \cite{ladder}.

The applied current can have a DC component (the so-called bias) and an AC
driving part.  The problem we are interested in is the behavior of the above
chain initially at rest and whose first junction only is submitted to AC
driving at a frequency $\Omega$ in the gap and intensity $I$. Then the model
results as \eqref{JSG} for $n>1$ and the bias $J=I_0$ together with the
following relation for $n=1$:
\begin{equation}\label{first}
\ddot u_1+\gamma\dot u_1+\sin u_1=c^2(u_2-u_1)+I_0+I\sin\Omega t.\end{equation}

It is important to remark that the above equation can be equivalently written
as the {\em discrete Neumann condition}
\begin{equation}\label{irrad}c(u_1-u_0)=-\frac Ic\sin\Omega t,\end{equation}
when the system \eqref{JSG} is assumed to hold also for $n=1$. This remark
actually allows us to consider the continuous limit in order to determine the
prediction of the threshold of nonlinear supratransmission.

\subsection{Bifurcation threshold prediction}

The continuous version of \eqref{JSG} and boundary value \eqref{irrad},
for the variable $x=n/c$ and $c\to\infty$, reads as the system
\begin{align}
& u_{tt}+\gamma u_t+\sin u=I_0+u_{xx},\label{JSG-cont}\\
& \left.\partial_x u\right|_{x=0}=B\sin\Omega t,\label{der-cont}
 \end{align}
where we have defined $B=-I/c$. This continous version corresponds to a long
Josephson junction whose extremity $x=0$ is submitted to external micro-wave 
irradiation, the amplitude $B$ being then related to the external magnetic 
field intensity \cite{antenna}. This is a Neumann boundary condition for the 
sine-Gordon continuous equation \eqref{JSG-cont}.

The system will then adapt the breather {\em derivative} at the boundary
centered in $-x_0$, namely
\begin{equation}\label{breath-deriv}
\left.\partial_x u_b\right|_{x=0}=4\frac{\kappa^2}\Omega
\frac{\sin\Omega t\ \sinh \kappa x_0}{\cosh^2\kappa x_0+
(\kappa^2/\Omega^2)\sin^2\Omega t},
\end{equation}
where now the continuous version of the dispersion relation is
$\Omega^2+\kappa^2=1$ for evanescent waves (due to the change of space
variable, we have $\kappa=c\lambda$ where $\lambda$ is defined in
\eqref{lambda}).

Upon varying the position $x_0$, the above expression has a maximum value
for $x_0=x_m$ given by $\sinh^2\kappa x_m=1+\kappa^2/\Omega^2$ and the
related maximum amplitude of the derivative eventually results as the
simple expression
\begin{equation}\label{threshold-deriv}
B_s=2(1-\Omega^2).\end{equation}
This is the threshold prediction for the Neumann condition \eqref{der-cont}
and we are going now to check it on numerical simulations.

\begin{figure}[ht]
\centerline{\epsfig{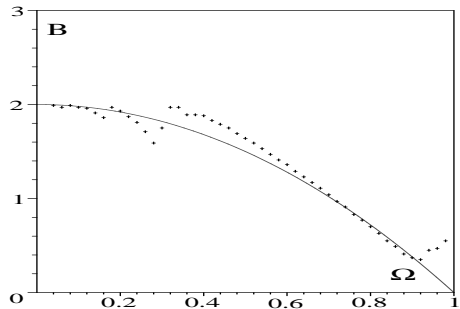}}
\caption{ Bifurcation diagram in the $(B,\Omega)$ plane. The solid line is the 
threshold expression \eqref{threshold-deriv}. The crosses indicate the lowest 
value of $B$ for which nonlinear supratransmission occurs.}
\label{fig:bif-JJ}\end{figure}
As in section \ref{sec:sg}, we solve the discrete system \eqref{JSG} with
no constant bias ($I=0$), without damping in a first stage ($\gamma=0$), for a
chain of $60$ particles with an absorbing boundary on the last $30$, and with a
coupling factor $c^2=25$. The result is displayed on figure \ref{fig:bif-JJ}
which shows good agreement except maybe around the frequencies $0.33$ and
$0.18$ as in figure \ref{fig:bif} (for the same reasons) and close to the
phonon band, which is due to the absorbing end.

\subsection{Josephson transmission line (JTL)}

Here above the situation is that of a quasi-continuous undamped and
semi-infinite sine-Gordon chain submitted to Neumann condition \eqref{der-cont}
a the origin. We turn back now to the discrete case, a one-dimensional 
finite-length (open-ended) array of coupled short Josephson junctions, where
the first pendulum is submitted to an external AC driving.  

Namely we consider the {\em Josephson transmission line} system
\begin{align}
&\ddot u_1+\gamma\dot u_1+\sin u_1=
c^2(u_2-u_1)+I_0+I\sin\Omega t,\label{forced}\\
&\ddot u_n+\gamma\dot u_n+\sin u_n=
I_0+c^2(u_{n+1}-2u_n+u_{n-1}),\label{JTL}\\
&u_{N+1}=u_N.\label{free}
\end{align}
Nonlinear supratransmission becomes here the property of the JTL to transmit
energy under the form of kinks (or anti-kinks) as soon as the intensity
$I$ of the AC driving of the first junction exceeds the threshold
\begin{equation}\label{thresh-JJ}
I_s=2c(1-\Omega^2)\end{equation}
as given by the definition $I=-cB$ and from expression \eqref{threshold-deriv}.

A new fact here is that, for Neumann type boundary conditions as \eqref{irrad},
only kinks (or anti-kinks) are produced, not breathers. This is the result of
the fact that producing a kink cost half of the energy used to produce a
breather, and of the freedom left on the boundary value $u_0(t)$ (only the
difference $u_1 - u_0$ is prescribed) allowing full $2\pi$-rotations. For
Dirichlet boundary condition as in preceding sections, the precription of
$u_0(t)$, e.g.  by \eqref{forcing} prevents full rotation. 

\begin{figure}[ht]
\centerline{\epsfig{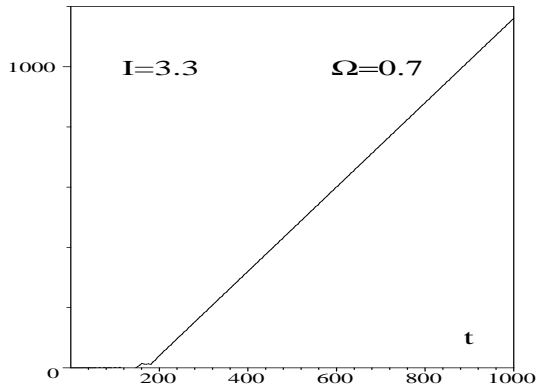}}
\caption{Motion of the particle $5$ in a JTL of 10 junctions with parameters
\eqref{simul-JJ}.}
\label{fig:out-kink}\end{figure}

To illustrate this property we draw in figure \ref{fig:out-kink} a typical
simulation with the following set of parameters
\begin{equation}\label{simul-JJ}
N=10\ ,\quad \gamma=0.1\ ,\quad c=3 \ ,\quad I_0=0.1\ ,
\end{equation}
constituting a reasonable choice for an experimental situation. At frequency
$\Omega=0.7$, the threshold \eqref{thresh-JJ} is $I_s=3.06$ and for an
AC driving at amplitude $I=3.3$ the chain starts to {\em rotate},
which means that a number of kinks (elementary $2\pi$ rotation) are
generated by the AC driving. In the case of figure \ref{fig:out-kink} about
$184$ kinks have been generated during $1000$ units of normalized time.

Note that we have assumed a small constant bias DC current
$I_0=0.1$ in order to select kinks instead of possible anti-kinks. However the
process does work without bias, though in some of the experiments the rotation 
would  stop after some time (for a reason that we do not understand). 
Finally we mention that we have selected compatible boundary condition,
namely, first an AC forcing that starts from zero and slowly reaches the value
$I$ within $100$ units of time, second the following initial positions and
velocities
\begin{equation}
u_n(0)=\arcsin(I_0)\ ,\quad \dot u_n(0)=0.\end{equation}
Then the nonlinear supratransmission does work in such JTL, despite the
small number of oscillators, the presence of damping, the free-end
boundary condition and the presence of constant bias. 

\subsection{Energy transmission}

\begin{figure}[ht]
\centerline{\epsfig{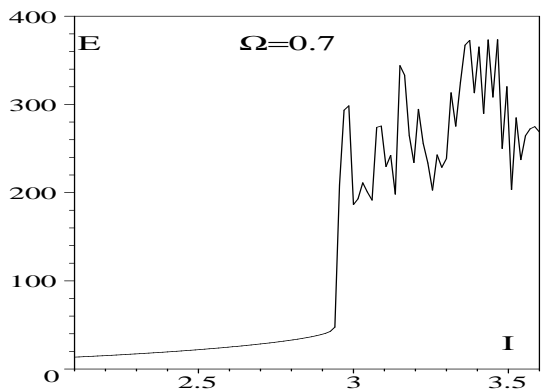}}
\caption{Energy transmitted for $180$ time units by a JTL  in terms of 
the driving intensity $I$.}\label{fig:JTL-ener}\end{figure}

In order to compute the transmitted energy in the present situation, we
start with expression \eqref{ener-tot} but written for a {\em finite number}
$N$ of oscillators, i.e.
\begin{equation}
E=\sum_{n=1}^{N}H_n+\frac{c^2}2(u_1-u_0)^2.\end{equation}
As before, this expression is differentiated and, by use of the conservation 
law \eqref{cons-law}, the free boundary condition $u_{N+1}=u_N$ leads to the
same result, namely
\begin{equation}
\frac{\partial}{\partial t}E=c^2\dot u_0(u_0-u_1).\end{equation}
Now, to compute the total energy injected in the medium during time $T$,
it is necessary to use integration by parts together with the Neumann
condition \eqref{irrad} and to chose for $T$ an integer multiple of the
period of the driving. We eventually obtain ($\ell=20$ in the
numerical simulations)
\begin{equation}\label{ener-JTL}
T=\ell \frac{2\pi}\omega\ ,\quad
E=-I\Omega\int_0^Tdt\ u_1(t)\cos\Omega t\ .\end{equation}
This is the quantity that we have computed, for each value of the
AC-driving $I$, in figure \ref{fig:JTL-ener}.

The bifurcation of energy translission is now quite clear on the graph of the
energy $E(T)$ transmitted to the chain displayed in figure \ref{fig:JTL-ener}
obtained for $\Omega=0.7$ and the parameters given in \eqref{simul-JJ}.

\section{Conclusion}

The ability of a nonlinear medium to transmit energy when submitted to periodic
boundary driving at a frequency in a stop gap and at an amplitude beyond a
threshold value, which we called nonlinear supratransmission, has been shown to
have some universality (so far in the domain of nonlinear wave type of
equations). The process does not rely on integrability and is robust against
damping, discreteness (in a reasonable range), finiteness and different classes
of boundary values. The mechanism at the origin of this process is a nonlinear
instability which is still under study.

Moreover, the nonlinear supratransmission has a simple experimental realization
in the pendula chain which works surprisingly well. Some other experimental
results are expected in Josephson transmission lines.

\paragraph*{Acknowledgements.} It is a pleasure to acknowledge enlighting
discussions with M.J. Ablowitz and M. Remoissenet,  A.V.  Ustinov, and decisive
experimental help of N.  Clementin, O. Guille, P. Munier and Y. Patin.

\end{document}